\documentclass[12pt]{article}
\usepackage{amsmath}
\usepackage{amsfonts}

\newtheorem{theo}{Theorem}

\def\0{{\bf 0}}

\def\R{\mathbb{R}}

\newcommand{\tod}{\stackrel{{\cal D}}{\longrightarrow}}

\newcommand{\eqco}{\setcounter{equation}{0}}
\newcommand{\thco}{\setcounter{theo}{0}}
\newcommand{\prco}{\setcounter{prop}{0}}
\newcommand{\laco}{\setcounter{lemm}{0}}
\newcommand{\coco}{\setcounter{coro}{0}}
\newcommand{\cjco}{\setcounter{conj}{0}}

\newcommand{\deco}{\setcounter{defn}{0}}
\newcommand{\allco}{\eqco  \thco \prco \laco \coco \cjco \deco}
\newcommand{\Var}{{\rm Var}}

\newcommand{\X}{{\cal X}}

\def\bdm{\begin{displaymath}}
\newcommand{\edm}{\end{displaymath}}
\def\benu{\begin{enumerate}}
\def\eenu{\end{enumerate}}
\def\beqn{\begin{equation}}
\def\eeqn{\end{equation}}
\def\be{\begin{equation}}
\def\ee{\end{equation}}
\def\bea{\begin{eqnarray}}
\def\eea{\end{eqnarray}}

\newcommand{\bean}{\begin{eqnarray*}}
\newcommand{\eean}{\end{eqnarray*}}
\newcommand{\bear}{\begin{eqnarray}}
\newcommand{\eear}{\end{eqnarray}}

\renewcommand{\epsilon}{\varepsilon}

\begin{document}

\title{\bf Mathematics of Random Growing Interfaces}

\author{Mathew D. Penrose and J. E. Yukich$^{1}$ \\
\\
{\normalsize{\em University of Durham and Lehigh University}} }
\date{}
\maketitle

\footnotetext[1]
{ Department of Mathematical Sciences, University
of Durham, South Road, Durham DH1 3LE, England: {\texttt
mathew.penrose@durham.ac.uk} }

\footnotetext[2]{ Department of Mathematics, Lehigh University,
Bethlehem PA 18015, USA: {\texttt joseph.yukich@lehigh.edu} }

\footnotetext{$~^1$ Research supported in part by NSA grant
MDA904-01-1-0029 }

\begin{abstract}
We establish a thermodynamic limit and Gaussian fluctuations for
the height and surface width of the random interface formed by
the deposition of particles on surfaces.  The results hold for the
standard ballistic deposition model as well as the surface
relaxation model in the  off-lattice setting. The results are
proved with the aid of general limit theorems for stabilizing
functionals of marked Poisson point processes.

\end{abstract}

\section{Introduction}
\allco

Aggregation and growth processes associated with the deposition
of particles on surfaces is of considerable scientific and
practical interest (Barab\'asi and H. E. Stanley \cite{BS},
Cumberland and  Crawford \cite{CC}, Vicsek \cite{Vik}).  Of
special interest is the morphology of the random interface
separating the growing material from its environment. The surface
morphology, which involves non-equilibrium growth behavior,
exhibits scaling and is thought to evolve to a steady state
without a characteristic time or length scale (Family \cite{Fa},
Family and Vicsek \cite{FV}).

This note rigorously develops the mathematical limit theory of the
random particle deposition model in which random size
$(d+1)$-dimensional balls
(`particles')  rain down sequentially at random onto a $d$
dimensional substrate $Q_n$ of volume $n^{d}$.  The substrate
$Q_n$  is assumed  initially flat but it need not be oriented
perpendicularly to the particle trajectories. Assume that
particle sizes are independent and identically distributed with
diameter bounded by a constant $D_1$.

When a particle arrives on the existing agglomeration of
deposited balls, it may stick to the first particle it contacts,
which may result in  lateral growth and `overhangs' (the
ballistic deposition (BD) model).  Alternatively, the particle may
slip and roll over existing particles, undergoing vertical and
horizontal displacements, stopping when it reaches a position of
lower height (the surface relaxation (SR) model).  In this model,
the particle diffuses along the interface up to a finite
distance,  say $D_2$ (c.f. \cite{Fa,BS}). Also, only
those existing particles within distance $D_1 + D_2$ of the new
particle affect its motion.   The finite displacement assumption
is consistent with the presence of moderately sticky particles.

Both the BD and SR models have been intensely studied in the
lattice and off-lattice (continuum) settings  (see \cite{BS} for
an overview of the enormous literature).  We focus here on the
continuum setting.

Let $\X \subset \R^d \times [0, \tau]$ denote a finite point set
of  particles, which are released from `infinity'
onto the substrate $\R^d$. Represent points in $\X$ by $(X,t)$,
where $X \in \R^{d}$ denotes the spatial location (center) of the
incoming particle and $t$ its time of arrival. The parameter
$\tau$ is the `mean deposition intensity', i.e., the average
number of particles arriving per unit volume in $\R^d$.


The `active zone' or the `interface' of the
resulting agglomeration  formed through either the BD or SR
process is defined as follows.
A  particle is termed
`exposed' if there is a positive chance that a subsequent incoming particle
could strike it
 directly.
  The active zone is the set $A(\X)$  of exposed
particles belonging to the agglomeration induced by $\X$.

Given $\X:=(X_1,...,X_k) \subset \R^d \times [0, \tau]$, let
$X_i^{'}, \ 1 \leq i \le k$, denote the position  of $X_i$ after
displacement.  Let $h(X'_i):=h(\X,X'_i)$ denote the height of
$X'_i $ above the substrate $\R^{d}$. Since the maximum diameter
is $D_1$ and the maximum displacement is $D_2$, the height $h(X)$
of a particle $(X,t)$ is a random variable determined by points
in the existing configuration within  a distance $D_1 +D_2$ of
$X$.

  Define the height
functional
\begin{equation} \label{H}
H(\X):= \sum_{X'_i \in A(\X) } h(X'_i), \ \ \X \subset \R^{d}
\times [0, \tau]
\end{equation}
and the mean height functional
\begin{equation} \label{Hbar}
\overline{H}(\X) := \frac{H(\X)} {\vert A(\X) \vert}, \ \ \X
\subset \R^d \times [0, \tau].
\end{equation}

The squared width of $X_i \in \X$ about the mean is
\begin{equation} \label{w}
w(X'_i):=w(\X, X'_i):=( h(X'_i) - \overline{H}(\X ) )^{2}, \ \ \X
\subset \R^d \times [0, \tau].
\end{equation}

The width functional is
\begin{equation} \label{W}
W(\X) := \sum_{X'_i \in A(\X) } w(X'_i), \ \ \X \subset \R^{d}
\times [0, \tau]
\end{equation}
and the surface width functional is
\begin{equation} \label{Wbar}
\overline{W} := \overline{W} (\X) :=
 \left( \frac{W(\X)} {\vert A(\X) \vert}
\right)^{1/2}, \ \ \X \subset \R^d \times [0, \tau],
\end{equation}
i.e., $\overline{W}$ is the root mean square of fluctuations of
heights of particles in the active zone.

We are interested in the mean height and surface width functionals
when $\X$ is the realization of a rate one homogeneous space time
Poisson point process ${\cal P}_{n, \tau}$ on $Q_n \times [0,
\tau], \ Q_n \subset \R^{d}$.  In this case we write $A(n,
\tau):= A({\cal P}_{n, \tau})$ for the active zone, and do not
distinguish between the active zones for the BD and SR models.
Let $ \{P_i\} $ denote the points in ${\cal P}_{n, \tau}$.

Define the {\em random height functional}
$H(n, \tau):= H( {\cal P}_{n, \tau})$
and the {\em mean random height functional}
$ \overline{H}(n, \tau) := \overline{H}({\cal P}_{n, \tau})
$. Similarly, define
the {\em random width functional} $W(n, \tau):= W({\cal P}_{n,
\tau})$ and the {\em random surface width functional}
$\overline{W}(n, \tau):= \overline{W} ({\cal P}_{n, \tau})$.


Considerable research has focused on the qualitative properties
of the random surface width $\overline{W}(n, \tau)$.  Its analysis is
rendered difficult by the presence of spatially correlated squared
heights $(h(P'_i) - \overline{H} )^2, \ P'_i \in A(n, \tau).$
Both the random height and random width functionals involve
triangular arrays of identically distributed but spatially
correlated summands.

 Beginning
with the seminal papers \cite{Fa}, \cite{FV}, and Kardar, Parisi,
and Zhang \cite{KPZ},
it is widely held that the surface width exhibits both space and
time scaling and is governed by the dynamic scaling relation
\begin{equation} \label{S1}
\overline{W}(n, \tau) \approx n^{\alpha} f(\tau/n^{\alpha/\beta} )
\end{equation}
where the scaling function $f$ satisfies
\begin{equation} \label{S2}
f(x) \approx  x^{\beta}, \ \ \ x \ll 1,
\end{equation}
and $f(x)\approx  C$  for $x \gg 1$. Here, $\alpha$ is called the
roughness exponent \cite{BS} and $\beta$ is the growth exponent;
$\alpha$ describes the dependence of the surface width in the
long time limit $\tau \to \infty$ on the linear size $n$, whereas
$\beta$ reveals the dependence on time $\tau$.  In the region $
\tau \ll n$, we have
\begin{equation} \label{S3}
\overline{W}(n, \tau) \approx \tau^{\beta}
\end{equation}
whereas for $\tau \gg n$,
\begin{equation} \label{S4}
\overline{W}(n, \tau) \approx n^{\alpha}.
\end{equation}

For both the BD and SD models there is a plethora of experimental
and theoretical results validating the phenomenological
expressions (\ref{S1}) - (\ref{S4}).  For an overview of the huge
literature, see \cite{BS}, \cite{Vik}.

Stochastic PDEs are  central to the analytic treatment of
continuum deposition models and the attendant phenomenological
surface width relations (\ref{S1}) - (\ref{S4}).  The classic
linear differential equation of
 Edwards and Wilkinson
 \cite{EW},
which treats the SR model, provides evidence for $\alpha = 1/2$
and $\beta = 1/4$ in dimension $d = 1$. The landmark paper
 \cite{KPZ} (KPZ) treats the BD model
and accounts for lateral growth by adding non-linear terms.
Subject to the validity of the KPZ methods, which are justified
by plausibility arguments and which cannot be formally derived
(see e.g. p. 56 of \cite{BS}), one obtains in dimension $d = 1$
the `exact' values $\alpha = 1/2$ and $\beta = 1/3$.  Higher
dimensional values of $\alpha$ and $\beta$ are not known.

Despite the abundance of experimental results and despite the
successes of PDEs, there is no formal, mathematically rigorous
treatment of the stochastic properties of the surface width for
the BD and SR models.

In this note we address this lack of mathematical precision by
providing a formal and rigorous probabilistic treatment of the
mean height and surface width functionals,
 establishing  a thermodynamic limit and Gaussian
fluctuations for the mean height and surface width functionals.
The mathematical underpinnings reside in the general limit theory
of stabilizing functionals of marked Poisson point processes
\cite{PY2}.  Although we do not obtain mathematical values for
the roughness and growth exponents,  the present approach provides
additional information on $\alpha$.   Attention here is confined
to the continuum, but the results also hold in the lattice
setting.

\section{Statement of Results}
\allco
\subsection{Thermodynamic limits}

Denote by ${\cal P}_{\tau}$ a rate one homogeneous Poisson point
process on $\R^{d} \times [0, \tau]$.  Without loss of
generality, for all $n = 1, 2,...$ we  let ${\cal P}_{n, \tau}$
be the restriction of ${\cal P}_{\tau}$ to $Q_n \times [0, \tau].$

Intuitively, one expects that the large $n$ limit of each summand
in the random height functional $H(n, \tau) := \sum_{P'_i \in
A(n, \tau) } h(P'_i)$ is related to a height functional defined
on the `infinite interface' created by the deposition of
particles in ${\cal P}_{\tau}$ on the infinite substrate
$\R^{d}$. However, it is not  straightforward
 to make rigorous sense of the notion
of sequential deposition of an infinite number of particles over a
substrate of infinite extent.

We adjoin to ${\cal P}_{\tau}$ a point $\{\0 \}  := \{(0,t) \}$,
with spatial coordinate  the origin of $\R^{d}$ and time
coordinate uniformly distributed over $[0, \tau]$.  The
collection of points ${\cal P}_{\tau} \cup \ \{\0 \} $ forms a
Palm distribution and we use it to define  `typical' height and
squared width fluctuation functionals.  Let ${\0}'  \in \R^{d}
\times [0, \tau]$ be the random position of $\{\0 \} $ after
displacement with respect to the incoming particles ${\cal P}_{n,
\tau}$, $h({\0}') := h({\cal P}_{n, \tau} \cup \{\0 \}, {\0}' )$
its random height,
and $A(n,\tau, 0)$ the active zone generated by ${\cal P}_{n,
\tau} \cup \{{\bf 0} \}$.

The following thermodynamic limits hold for both the BD and SR
models.
  They involve notions  of {\em conditional
expectation} and {\em conditional variance}.
Given a  a random variable $\zeta$ and an event $B$ with non-zero
probability, the conditional distribution of $\zeta$ given
$B$ is the probability distribution $P[\{\zeta \in \cdot \} \cap B]/P[B]$;
the mean and variance of this conditional distribution
are denoted $E[\zeta|B]$ and ${\rm Var}[\zeta|B]$.

\begin{theo} \label{LLN1} (a) For all $\tau \in (0, \infty)$, the
random  limits
 $$ h_0(\tau)
:= \lim_{n \to \infty} h( {\0}' ) \cdot
1_{ \{ {\0}'  \in A( n,\tau, 0) \} } \ \text{and}  \ \ \
\xi_0(\tau) := \lim_{n \to \infty}  1_{ \{ {\0}'  \in A( n,\tau,
0) \} }
$$
exist a.s.

(b) For all $\tau \in (0,\infty)$,
$$
\lim_{n \to \infty} \frac{\vert A(n, \tau) \vert} {n^d } = \tau
\cdot E[\xi_0(\tau) ] \ \ a.s.
$$

(c) For all $\tau \in (0, \infty)$,
$$
\lim_{n \to \infty} \overline{H}({\cal P}_{n, \tau})
= \frac{ E[h_0 (\tau) ] } { E[\xi_0(\tau) ] }
= E[h_0(\tau) \vert \xi_0(\tau) = 1 ]  \ \ a.s.
$$
\end{theo}

If we interpret $h(  {\0}' ) \cdot 1_{ \{ {\0}'  \in A( n,\tau,
0)\} }$ as the  height of a `typical' point in the interface,
then the `typical' squared width (height fluctuation) of a point
in the growing interface $A(n, \tau, 0)$ is given by
\begin{equation}
w( {\0}' ) \cdot 1_{ \{ {\0}'  \in A( n,\tau, 0) \} } := w({\cal
P}_{n, \tau} \cup \{\0 \}, {\0}' ) \cdot 1_{ \{ {\0}' \in A(
n,\tau, 0) \} }.
\end{equation}

The next result shows that the typical squared width a.s.
converges in the large $n$ limit, and moreover the limit governs
the fluctuations of the interface width.

\begin{theo} \label{LLN2} (a) For all $\tau \in (0, \infty)$, the
limit $$  w_0(\tau) := \lim_{n \to \infty} w( {\0}' ) \cdot 1_{
\{ {\0}'  \in A( n,\tau, 0) \} }
$$
exists a.s.

(b) For all $\tau \in (0, \infty)$, \bean \lim_{n \to \infty}
\overline{W}^2({\cal P}_{n, \tau}) := \lim_{n \to \infty}
\frac{1}{\vert A(n, \tau) \vert  } \sum_{P'_i \in A(n,\tau) }
w(P'_i) = \frac{E[h^2_0(\tau)] } {E[ \xi_0(\tau) ] } - \left(
\frac{ E[h_0(\tau) ] } {E[\xi_0(\tau) ] } \right)^2
\\
= {\rm Var}[h_0(\tau) \vert \xi_0(\tau)= 1] \\
= E[w_0(\tau) \vert \xi_0(\tau) = 1 ]  \ \ a.s. \eean
\end{theo}

{\em Remarks}

(i) Just as the random variable $h_0(\tau)$ gives meaning to the
height of a `typical' particle in an infinite interface defined
by a Poisson point process, $w_0(\tau)$ may be interpreted as a
`typical' width.  Similarly, $E[\xi_0(\tau) ]$ is the probability
that a typical particle belongs to the active zone.

(ii) The scientific literature assumes the existence of the limits
$\lim_{n \to \infty} \overline{H}(n, \tau)$ and $\lim_{n \to
\infty} \overline{W}^2(n, \tau)$.   Theorems \ref{LLN1} and
\ref{LLN2} puts this on solid footing and identifies the limiting
constants.
 In statistical terms, Theorems
\ref{LLN1} (b) and \ref{LLN2} (c) say that  the sample mean
(variance) of the heights of the particles in the interface form
a consistent
 estimator for the mean (variance) of the random height of
a typical particle, conditional on its being in the interface
at all, as the substrate becomes large.

(iii)  For both the BD and SR models, it is of great interest to
 estimate the constants in Theorems \ref{LLN1}(c) and
\ref{LLN2}(b).
 Heuristically, one expects that
$E[h_0 (\tau) |\xi_0(\tau) =1 ]  \approx C \cdot \tau$
  since the expected height should
vary directly with the deposition intensity $\tau$.
  For large
$n$, we expect that $ {\0}' \in A(n, \tau, 0) $ with a probability
inversely proportional to $\tau$, i.e., we expect $E[\xi_0 (\tau)
] \approx 1/\tau.$ Also, given $\xi_0(\tau) = 1$, one expects that
the expectation of the squared width of a `typical' particle in
the {\em smoothed} interface, namely
$E[w_0(\tau) \vert \xi_0(\tau)= 1]$ is upper bounded by the
expectation of the width of a `typical' particle in the {\em
non-smoothed} interface. The latter is just the variance of a
Poisson random variable with parameter $\tau$, giving
$E[w_0(\tau) \vert \xi_0(\tau) = 1] = O(\tau)$, which is
consistent with the simulations of Zabolitzky and Stauffer
(\cite{ZS}, p. 1529). Whether this bound can be improved is the
subject of widespread research.  A rigorous mathematical
treatment of $E[w_0(\tau) \vert \xi_0(\tau) = 1]$ would be
welcome.

\subsection{Gaussian fluctuations}

Both the height functional $H$ and a modified version of surface
width exhibit Gaussian fluctuations about the mean, as shown by
the following distributional results. $N(0,\sigma^2)$ denotes a
normal random variable with mean $0$ and variance $\sigma^2$,
and $\tod$ denotes convergence in distribution.

\begin{theo} \label{CLT1}  For all $\tau \in (0, \infty)$ there
exists $\sigma_{\tau}^2:= \sigma^2_{\tau}(H)  > 0$ such that
$$
\frac{ H({\cal P}_{n, \tau}) - E H({\cal P}_{n, \tau})} {n^{d/2}
} \tod N(0, \sigma_{\tau}^2 )
$$
and $n^{-d} \Var H({\cal P}_{n, \tau}) \to \sigma_{\tau}^2.  $
\end{theo}

Let $D(\tau):= \frac{ E[h_0 (\tau) ] } { E[\xi_0(\tau) ] }$
denote the mean deposition height given by Theorem
\ref{LLN1}(c).  The following is closely related to the surface
width functional
\begin{equation} \label{modW}
W_1({\cal P}_{n, \tau}):= W_1(n, \tau):= \sum_{P'_i \in A(n,
\tau)} (h(P'_i) - D(\tau) )^2.
\end{equation}

The  modified width functional $W_1$ exhibits Gaussian
fluctuations around its mean in the large $n$ limit.

\begin{theo} \label{CLT2}  For all $\tau \in (0, \infty)$ there
exists $\sigma_{\tau}^2:= \sigma^2_{\tau}(W_1)  > 0$ such that
$$
\frac{ W_1 ({\cal P}_{n, \tau}) - E W_1 ({\cal P}_{n, \tau})}
{n^{d/2} } \tod N(0, \sigma_{\tau}^2 )
$$
and $n^{-d}\Var W_1 ({\cal P}_{n, \tau})  \to \sigma_{\tau}^2. $
\end{theo}

{\em Remarks}

(i) If the Poisson point process ${\cal P}_{n, \tau}$ is replaced
by the binomial point process $\{U_{n,1},...,U_{n, [n \tau]} \}$,
where $U_{n,1},U_{n,2},...$ are i.i.d. with the uniform
distribution on $Q_n \times [0, \tau]$, then Theorems \ref{LLN1}
and \ref{LLN2} continue to hold.  Theorems \ref{CLT1} and
\ref{CLT2} also remain valid, albeit with
smaller variances $\sigma^2_{\tau}(H)$ and $\sigma^2_{\tau}(W_1)$.
 See \cite{PY2} for details.

(ii) Theorems \ref{LLN1}-\ref{CLT2} hold for particles of random
shapes and types;  the particles need not be balls \cite{PY2}.
The substrate $Q_n$ need not be a cube, but may be a more general
region subject to a smoothness assumption on its boundary
\cite{PY2}.


(iii) We have not been  able to determine the dependence of
$\sigma^2_{\tau}(H)$ and $\sigma^2_{\tau}(W_1)$ on $\tau$. The
exact order of magnitude of $\sigma^2_{\tau}(H)$ and
$\sigma^2_{\tau}(W_1)$ could help determine the scaling exponents
$\alpha$ and $\beta$ in (\ref{S1}) - (\ref{S3}).

(iv) Extending Theorems \ref{LLN1}-\ref{CLT2} to cases involving
$\tau$ and $n$ going to infinity together appears beyond the
scope of the methods here. Such cases are of interest and merit
investigation, particularly since they might also help determine
values for $\alpha$ and $\beta$.

\section{Methods}
\allco


To prove limit results for the mean height and width functionals,
we write
\begin{equation}
\label{apr2}
 \overline{H}(n, \tau) := \overline{H}({\cal P}_{n, \tau}):=
 \frac{n^{d} } {\vert A(n, \tau) \vert } \frac{ \sum_{P'_i \in A(n,\tau)}
  h(P'_i) } {n^{d} }
\end{equation}
and
\begin{equation}
 \overline{W}^2(n, \tau) := \overline{W}^2({\cal P}_{n, \tau}):=
 \frac{n^{d} } {\vert A(n, \tau) \vert } \frac{ \sum_{P'_i \in A(n,\tau)}
  w(P'_i) } {n^{d} }.
\end{equation}

To prove a thermodynamic limit for $\overline{H}({\cal P}_{n,
\tau})$
it will suffice to show that the  quotients $\frac{n^{d} } {\vert
A(n, \tau) \vert }$ and $\frac{ \sum_{P'_i \in A(n,\tau)}
  h(P'_i) } {n^{d} }\ $   both tend to a limit a.s.  The same
  is true for $\overline{W}({\cal P}_{n, \tau})$. Notice that
  we may write
  \begin{equation}
| A(n, \tau) | := \sum_{P_i \in {\cal P}_{n, \tau} }  \xi({\cal P}_{n,
\tau},P_i),
\end{equation}
where $\xi(\X, x) = 1$ or $0$, depending upon whether $x$ is
exposed or not in the agglomeration induced by $\X$.

To extract  limits for the height and width functionals, as well
as for the number of exposed points, we probe their behavior by
adjoining $\{{\bf 0} \}$ to $\X$, where we recall that
 $\{{\bf 0} \}$ has as spatial coordinate  the origin of
$\R^{d}$ and a time coordinate  randomly distributed over $[0,
\tau]$. The change in the height functional (the `add one cost'
\cite{PY2}) caused by the insertion of $\{{\bf 0} \}$ is given by
\begin{equation} \label{addone}
\Delta_H(\X):=
 H(\X \cup \{{\bf 0} \})
- H(\X)
\end{equation}
with a similar definition for $\Delta_W(\X)$ and
$\Delta_{W_1}(\X).$

The displacement of the incoming particle $\{{\bf 0} \}$ is
clearly determined by the location and displacement of particles
previously arriving within a distance
$D_3 := 2( D_1 +  D_2)$ of the
origin of $\R^{d}$. Whether  the particle ends up in the active
zone $A(\X)$ is clearly a function of particles arriving within a
distance
$D_3$ of the origin.

Also, the insertion of the particle $\{{\bf 0} \}$ will only affect
a later arriving particle at location $x \in \R^{d}$ if there is
a sequence of points $\{(x_i, t_i)\}_{i=1}^n $ in $\X$ such that
$$
 \vert x_1 \vert \leq D_3 ,
$$
$$
 \vert x_i - x_{i+1} \vert \leq D_3, \ \ \text{and} \ \ t_i <
t_{i+1}, \ \ 1 \leq i \leq n -1,$$ and
$$
 \vert x_n - x \vert \leq D_3.
$$

When $\X$ is the homogeneous Poisson point process ${\cal P}_{n,
\tau}$, then such sequences exist with a probability decaying
exponentially in $\vert x \vert$ \cite{PY2}.
%
%
%
Hence, adjoining $\{{\bf 0} \}$ to ${\cal P}_{n, \tau}$ changes
the heights $h(P'_i)$ of the re-positioned particles $P'_i, P_i
\in {\cal P}_{n, \tau}$, $\vert P_i \vert \geq \lambda$, with a
probability which decays exponentially in $\lambda$.  (Thus
$h(P'_i)$  and $\xi(P'_i)$ have exponentially decaying
correlations.)
The insertion of $\{{\bf 0} \}$ thus changes the morphology of the
active zone at distances larger than $\lambda$ from the origin
with a probability decaying exponentially in $\lambda$.

We conclude that for all deposition intensities $\tau \in (0,
\infty),$ there is an a.s. finite random variable
$N:= N_H(\tau)$, with exponentially decaying tails, and an a.s.
finite random variable $\Delta(\tau):= \Delta_H(\tau)$ such that
with probability one
\begin{equation} \label{Hstable}
 H( ({\cal
P}_{\tau} \cap Q_N) \cup
 \{{\bf 0} \} \cup {\cal A} )
- H( ({\cal P}_{\tau} \cap Q_N) \cup {\cal A})
= \Delta(\tau)
 \end{equation}
for all finite ${\cal A} \subset (\R^d \setminus Q_N) \times [0, \tau]) $;
see \cite{PY2} for details.  The  `add one cost' to the
height functional is unaffected by changes in the particle
configuration outside  $Q_N \times [0, \tau]$. Following extant
terminology, \cite{PY2}, we say that the height functional
`stabilizes'.

Moreover, for the same reasons, $h$ also `stabilizes' in the
sense that there are a.s. finite random variables $N$ and
$h_0(\tau) $  such that
\begin{equation} \label{hstable}
h( ({\cal P}_{\tau} \cap Q_N ) \cup \{ \0 \} \cup {\cal A}, {\0}'
) \cdot
 1_{\{ \0' \in A( ({\cal P}_{\tau} \cap Q_N ) \ \cup
 \  \{\0\}  \ \cup {\cal A} )  \} }  = h_0(\tau)
 \end{equation}
for all finite
 ${\cal A} \subset (\R^{d} \setminus Q_N) \times [0, \tau]. $ This
last fact yields Theorem \ref{LLN1}(a).

Next, given a subcube $Q \subset \R^{d},$ let $U_{1,Q},
U_{2,Q},\ldots$ be i.i.d. uniformly distributed random variables on
$Q \times [0, \tau]$ and let ${\cal U}_{m,Q}:=
\{U_{1,Q},U_{2,Q},...,U_{m,Q} \}$ be a binomial point process.
Adjoin the particle $\{{\bf 0} \}$ to  ${\cal U}_{m,Q}$ and
consider the add one cost
\begin{equation} \label{addonecost}
\Delta_H({\cal U}_{m,Q}):=
  H({\cal U}_{m,Q} \cup \{ \0 \})
- H({\cal U}_{m,Q}).
\end{equation}
Since $H$ stabilizes and since $N_H(\tau)$ has finite  moments of
all orders \cite{PY2}, by coupling binomial and Poisson point
processes we can show that $\Delta_H({\cal U}_{m,Q})$ has finite
moments of all orders.  In fact, for all $\tau \in [0, \infty)$
and all $p > 0$, $H$ satisfies the bounded moments condition
\begin{equation} \label{bmc}
\sup_{Q } \sup_{m \in [\tau \vert Q \vert/2, \ 3 \tau \vert Q
\vert/2 ]} E [ \Delta_H({\cal U}_{m,Q})^p ] < \infty,
\end{equation}
where the first supremum runs over cubes $Q \subset \R^{d}$ containing the
origin \cite{PY2}.

Likewise,
 the functional $h$ satisfies the bounded second moments
condition
\begin{equation} \label{bmch}
\sup_{Q} \sup_{m \in [\tau \vert Q \vert/2, \ 3 \tau \vert Q
\vert/2 ]} E [ h( {\cal U}_{m,Q}  \cup  \{\0 \}, {\0}' )^2 ] <
\infty.
\end{equation}

By combining the translation invariance of $h$, the uniform  bound
$H(\X) \leq C_1 (\text{card} (\X) )^2$,
 the bounded moments conditions (\ref{bmc},
\ref{bmch}), and by invoking the stabilization of $h$ (\ref{hstable}),
one verifies the hypotheses of the general law of large numbers
given by Theorem 3.2 of \cite{PY2}.  We thus deduce for all $\tau
\in (0, \infty)$,
\bea
\lim_{n
\to \infty} \frac{1}{n^{d} } \sum_{P'_i \in A(n,\tau) } h(P'_i) =
\tau \cdot E[h_0 (\tau) ] \ \ \ a.s.
\label{mar30}
\eea

Similarly, using the stabilization of $\xi$ and again using the
general law of large numbers of \cite{PY2},  we may show for all
$\tau \in (0, \infty)$ that $\lim_{n \to \infty} \frac{|A(n,
\tau)|} {n^{d}} = \tau \cdot E[\xi_0(\tau) ]$. This gives Theorem
\ref{LLN1} (b). Combining this last limit with (\ref{mar30}) and
(\ref{apr2}), we obtain Theorem \ref{LLN1}(c).

By invoking the stabilization of $H$, we may apply the general
central limit theorem (Theorem 3.1) of \cite{PY2} to deduce
Theorem \ref{CLT1}. According to Theorem 3.1 of \cite{PY2},  the
limiting variance $\sigma_{\tau}^2(H)$ is non-zero provided that
$\Delta_H(\tau)$ is non-degenerate. To check non-degeneracy of
$\Delta_H(\tau)$, observe that there is positive probability that
the set ${\cal P}_{n, \tau}$ puts no particles within distance
$D_3$ of the origin, so that in this case the difference
\begin{equation} \label{diff}
 H( ({\cal
P}_{\tau} \cap Q_N) \cup \{ \0 \} \cup {\cal A} )
- H( ({\cal P}_{\tau} \cap Q_N) \cup {\cal A})
 \end{equation}
is just the height of a single particle resting on the substrate.
On the other hand, there is also positive probability that the set
${\cal P}_{n, \tau}$ is such that the addition of the origin
results in a net increase of one exposed point to the active
surface, in such a way that the height is increased by the height
of a particle which is several layers removed from the substrate.
This produces a different value for (\ref{diff}), showing the
non-degeneracy of $\Delta_H(\tau)$.

  We prove Theorems \ref{LLN2} and \ref{CLT2} in a similar fashion.
Define the functionals
$$
w_1(X'):= w_1(\X,X'):= (h(X') - D(\tau) )^2
$$
and
$$
W_1(\X):= \sum_{X'_i \in A(\X)} w_1(\X,X'_i).
$$

Analogously to (\ref{hstable}), the functional
$$
w_1({\cal P}_{n, \tau} \cup
 \{ \0 \}, {\0}'):= (h({\0}' ) - D(\tau) )^2
 $$
stabilizes in the sense that for all $\tau \in (0, \infty)$ there
is a random variable  $N$ with exponentially decaying tails and
$w_0(\tau)$ a.s. finite
  such that
 $$
 w_1( ({\cal P}_{\tau} \cap Q_N ) \cup
 \{ \0 \} \cup {\cal A}, {\0}') \cdot 1_{ \{ \0'  \in A(n,\tau,0) \} }  = w_0(\tau)
 $$
 for all finite ${\cal A} \subset (\R^{d} \setminus Q_N) \times [0,\tau].$%

 Moreover, since $W_1$ satisfies the bounded  moments
 condition (\ref{bmc}) (replace $\Delta_H$ by $\Delta_{W_1}$
 there) and since $w_1$ satisfies the bounded second moments
 condition (\ref{bmch})  (replace $h$ by
 $w_1$ there), we obtain for all $\tau \in (0, \infty)$,
 \begin{equation}\label{limits}
 w_0(\tau) = \lim_{n \to \infty}w_1({\cal P}_{n, \tau} \cup
 \{ \0 \}, {\0}')\cdot 1_{ \{ {\0}'  \in A(n,\tau,0) \} } \ \ a.s.
 \end{equation}
 Since $\overline{H} \to D(\tau)$ a.s. by  Theorem
 \ref{LLN1}(c), it follows that
$$
 w_0(\tau) = \lim_{n \to \infty}w({\cal P}_{n, \tau} \cup
 \{ \0 \} , \0') \cdot 1_{ \{ \0'  \in A(n,\tau,0) \} } \ \ a.s.
 $$
 i.e., Theorem \ref{LLN2}(a) holds.

To prove Theorem \ref{LLN2}(b), we write
\begin{equation}\label{WW}
W({\cal P}_{n, \tau}) = \sum_{P'_i \in A(n, \tau) } \left(h(P'_i)
- \overline{H}(n,\tau) \right)^2 = \sum_{P'_i \in A(n, \tau)
}h^2(P'_i) -| A(n, \tau) | (\overline{H}(n, \tau) )^2.
\end{equation}


Dividing both sides of (\ref{WW}) by $A(n, \tau)$ and appealing to
Theorem \ref{LLN1}(c), we see that
 to
establish a thermodynamic limit for  $\overline{W}({\cal P}_{n,
\tau})$ it is enough to show a thermodynamic limit for $
\frac{\sum_{P'_i \in A(n, \tau) }h^2(P'_i) } {|A(n, \tau)|}$.
However, by following the proof of Theorem \ref{LLN1} with $h$
replaced by $h^2$, we find that
$$
\lim_{n \to \infty} \frac{\sum_{P'_i \in A(n, \tau) }h^2(P'_i) }
{|A(n, \tau)|} = \frac{ E[h^2_0 (\tau) ] } { E[\xi_0(\tau) ] }.
$$
Using Theorem \ref{LLN1}(c) on the second term in (\ref{WW}), we
obtain the first equality in Theorem \ref{LLN2}(b).

The second equality in Theorem \ref{LLN2}(b) is just the
definition of conditional variance and it remains to show the last
equality. Let $E_n:= { \{ \0' \in A(n,\tau,0) \} }$ for
simplicity. We have
 \bean & (h_0(\tau)
- D(\tau) )^2 \\
& = ( \lim_{n \to \infty}  h( {\0}') 1_{E_n} - D(\tau) )^2 \\
& = \lim_{n \to \infty} h^2( {\0}' ) \cdot 1_{E_n} - 2D(\tau)
\lim_{n \to
\infty} h( {\0}' ) 1_{E_n} + D(\tau)^2 \\
& = \lim_{n \to \infty} (h( {\0}' ) - D(\tau))^2 \cdot 1_{E_n} +
D(\tau)^2
\lim_{n \to \infty} 1_{E_n^c} \\
& = \lim_{n \to \infty} w_1({\cal P}_{n, \tau} \cup
 \{ \0\} , \0')  \cdot 1_{E_n} + D(\tau)^2
\lim_{n \to \infty} 1_{E_n^c} \\
& = w_0(\tau) + D(\tau)^2 (1 - \xi_0(\tau)),
 \eean
by (\ref{limits}) and by Theorem \ref{LLN1}(a). Thus, by the
above identity
 \bean {\rm
Var}[h_0(\tau) \vert \xi_0(\tau)= 1] = E[(h_0(\tau) - E[h_0(\tau)
\vert \xi_0(\tau) = 1])^2 \vert \xi_0(\tau) = 1 ]
\\
= E[(h_0(\tau) - D(\tau))^2 \vert \ \xi_0(\tau) = 1 ] =
E[w_0(\tau) \vert \xi_0(\tau) = 1], \eean
 showing the last equality in Theorem \ref{LLN2}(b).

Finally, we address the proof of Theorem \ref{CLT2}.
 Exactly as for the height functional $H$, we can show that $W_1$
 stabilizes in the sense that (\ref{Hstable}) holds with
 $H$ replaced by $W_1$. Invoking the stabilization
 of $W_1$ and applying Theorem 3.1 of
\cite{PY2}, we obtain Theorem \ref{CLT2}. Non-degeneracy of
$\Delta_{W_1}(\tau)$ follows as in the proof of non-degeneracy of
$\Delta_{H}(\tau)$.





\end{document}